\documentclass[12pt,a4paper]{article}

\usepackage{epsfig}
\usepackage{amsmath,amsfonts,amssymb}
\usepackage{t1enc}
\usepackage{cite}
\usepackage{colortbl}
\usepackage{pifont}

\providecommand{\openone}{\leavevmode\hbox{\small1\kern-3.8pt\normalsize1}}

\newcommand{\RE}{\text{Re}\,}

\newcommand{\bmu}{\mathcal{B}_\mu}
\newcommand{\wmu}{\mathcal{W}_\mu}
\newcommand{\gmu}{\mathcal{G}_\mu}
\newcommand{\hmu}{\mathcal{H}_\mu}
\newcommand{\qmu}{\mathcal{Q}_\mu^5}
\newcommand{\ymu}{\mathcal{Y}_\mu^5}
\newcommand{\Of}{\Omega^4}
\newcommand{\So}{\Sigma}

\newcommand{\oh}{\textstyle \frac{1}{2}}
\newcommand{\ot}{\textstyle \frac{1}{3}}
\newcommand{\of}{\textstyle \frac{1}{4}}
\newcommand{\os}{\textstyle \frac{1}{6}}

\newcommand{\otw}{\textstyle \frac{1}{12}}

\newcommand{\fsx}{\textstyle \frac{5}{6}}
\newcommand{\ssx}{\textstyle \frac{7}{6}}

\newcommand{\gM}{\gamma^\mu}
\newcommand{\gm}{\gamma_\mu}
\newcommand{\la}{\lambda^a}
\newcommand{\tI}{\tau^I}

\newcommand{\Cqq}{C_{qq}}
\newcommand{\Cqqp}{C_{qq'}}
\newcommand{\Cuu}{C_{uu}}
\newcommand{\Cqu}{C_{qu}}
\newcommand{\Cqup}{C_{qu'}}

\newcommand{\afb}{A_\text{FB}}

\newcommand{\twt}{\textstyle \frac{2}{3}}

\begin{document}

\begin{center}
\begin{Large}
{\bf No like-sign tops at Tevatron: \\
Constraints on extended models \\[3mm]
and implications for the $t \bar t$ asymmetry.}
\end{Large}

\vspace{0.5cm}
J. A. Aguilar--Saavedra, M. P\'erez-Victoria \\[0.2cm] 
{\it Departamento de F\'{\i}sica Te\'orica y del Cosmos and CAFPE, \\
Universidad de Granada, E-18071 Granada, Spain}
\end{center}

\begin{abstract}
We use a recent upper limit from CDF on like-sign top pair production to place constraints on general new vector bosons and scalars mediating $u u \to t t$. The possible vector bosons comprise neutral colour singlets or octets, and charge $4/3$ colour triplets or sextets, whereas the new scalars can be neutral colour singlets or octets and charge $4/3$ colour sextets. We also estimate the expected bounds from like-sign top pair production at LHC in the near future. Then, we address the implications of these limits for the forward-backward asymmetry in $t \bar t$ production measured at Tevatron. In particular, we find that models explaining the observed asymmetry by the exchange of a single $t$-channel heavy $Z'$ boson are already excluded. On the other hand, light $Z'$ bosons  with a mass $M_{Z'} \simeq 150$ GeV, which could also account for a recent CDF dijet excess in $W+$jet production, are barely allowed.
\end{abstract}

\section{Introduction}

The production of like-sign top quark pairs would be a striking signature of physics beyond the Standard Model (SM). At hadron colliders, charge conservation implies that $tt$ pairs can only be produced from initial up or charm quarks. Hence, proton-proton colliders are better suited for studying this signal, in particular in $uu \to tt$. In fact, like-sign top production is a golden channel for early discoveries at the Large Hadron Collider (LHC)~\cite{Berger:2010fy,Bauer:2009cc}. Maybe more surprising is the fact that the high statistics accummulated by Tevatron already allows to extract useful limits on various SM extensions mediating this process.

Recently, the CDF collaboration has set a limit on like-sign top production at Tevatron, using a luminosity of 6.1 fb$^{-1}$ \cite{CDFtt},
\begin{equation}
\sigma(tt+\bar t \bar t)\times \text{Br}(W \to \ell \nu)^2 < 54~\text{fb} \,,
\label{ec:lim}
\end{equation}
with a 95\% confidence level (CL).
As we shall show, this limit puts significant constraints on the different SM extensions that can potentially give observable contributions to $uu \to tt$. These extensions must contain boson fields mediating this process at tree level. Assuming renormalizable interactions, these fields can be either extra vector bosons or extra scalars. Since the new interactions must fulfill the $\text{SU}(3)_C\times \text{SU}(2)_L \times \text{U}(1)_Y$ gauge invariance of the SM, the quantum numbers of the new fields are not arbitrary. It is easy to check that the only possibilities for particles of spin 1 are~\cite{delAguila:2010mx}
\begin{itemize}
\item a neutral colour-singlet $Z'$, which can be in a singlet $\text{SU}(2)_L$ representation (denoted here as $\bmu$) or belong to a triplet $\wmu$;
\item a neutral colour-octet $g'$, either an isosinglet $\gmu$ or member of an isotriplet $\hmu$;
\item the charge $4/3$ component of a colour-triplet isodoublet $\qmu$;
\item the charge $4/3$ component of a colour-sextet isodoublet $\ymu$.
\end{itemize}
On the other hand, the possible new particles of spin 0 are
\begin{itemize}
\item the neutral scalar of an isodoublet, which can be a colour singlet $\phi$ or octet $\Phi$;
\item the charge $4/3$ component of a colour-sextet, either an isosinglet $\Of$ or included in an isotriplet $\So$.
\end{itemize}
Obviously, charge $4/3$ and neutral particles are exchanged in $s$ and $t$ channels, respectively. 
In this letter, we translate the CDF upper limit in Eq.~(\ref{ec:lim}) into constraints on the extended sectors enumerated above. Furthermore, we compare with potential LHC measurements with 2010 data (35 pb$^{-1}$) and 2011 data (1 fb$^{-1}$). In particular, we place direct limits on 
flavour-changing neutral (FCN) couplings of new neutral vector bosons $Z'$, $g'$ as well as neutral scalars. In order to unify the computations for the different fields, we make use of effective field theory in the following manner: we first integrate out the new heavy states and obtain their contribution to $uu \to tt$ in terms of four-fermion operators; then, we obtain the cross section in terms of effective operator coefficients. In an appendix we discuss the range of validity of this approximation and present exact results for light $Z'$ bosons.

We devote special attention to the implications of these limits for the $t \bar t$ forward-backward (FB) asymmetry at Tevatron. One mechanism that has been suggested to enhance this asymmetry and thus explain the measured values, in particular $\afb = 0.475 \pm 0.114$ for $m_{t\bar t} > 450$ GeV~\cite{Aaltonen:2011kc},
is the exchange of a flavour-violating $Z'$ boson in the $t$ channel~\cite{Jung:2009jz,Cao:2009uz,Choudhury:2010cd,
Cao:2011ew,Bhattacherjee:2011nr,Berger:2011ua}. It is well known that, for a single real $Z'$, this automatically implies like-sign $tt$ production. The relation between the two processes can easily be understood, without precise knowledge of the details to be given below, by taking the CP conjugate of one of the two vertices (the fact that the vector boson is real is crucial). Here, we show that the non-observation of like-sign tops at Tevatron already rules out these models as the sole explanation of the Tevatron $t \bar t$ asymmetry, except for very light $Z'$ masses which, interestingly, are consistent with a recent CDF dijet excess~\cite{Aaltonen:2011mk}. We also note that this direct relation between $tt$ production and the value of $\afb$ does not hold any longer when more than one $Z'$ boson is present, as in the model in Ref.~\cite{Jung:2011zv}. Among the other new particles that can enhance the FB asymmetry, some belong to the list above. Therefore, the limits from $tt$ production reduce the parameter space of extensions with these particles.

\section{New bosons and $tt$ production}
\label{sec:2}

The process $uu \to tt$ is absent in the SM at the tree level but it can be mediated by new vector bosons in six possible $\text{SU}(3)_C \times \text{SU}(2)_L \times \text{U}(1)_Y$ (irreducible) representations~\cite{delAguila:2010mx}, or by scalars in four possible representations.\footnote{Notice that for scalars mixing up-type quarks there are two additional $\text{SU}(3)_C$ triplet representations~\cite{AguilarSaavedra:2011vw}: 
isotriplets coupling to $q_{Li} q_{Li}^c$ and isosinglets coupling to $u_{Ri} u_{Rj}^c$. However, their coupling matrices are anti-symmetric and diagonal couplings to $uu^c$ and $tt^c$ vanish.} They are all collected in Table~\ref{tab:lagr}, where in the first column we write the symbol used to label them. The relevant interaction Lagrangian is included as well. In the last column we display the symmetry properties, if any, of the coupling matrices $g_{ij}$.
\begin{table}[p]
\begin{center}
\begin{tabular}{c|clc}
Symbol & Rep. & \multicolumn{1}{c}{Interaction Lagrangian} & Sym. \\
\hline
$\bmu$ & $(1,1)_0$ 
  & $-\left( g_{ij}^q \bar q_{Li} \gM q_{Lj} 
  + g_{ij}^u \bar u_{Ri} \gM u_{Rj} 
  + g_{ij}^d \bar d_{Ri} \gM d_{Rj} \right) \bmu $ & $g=g^\dagger$ \\[1mm]
$\wmu$ & $(1,\text{Adj})_0$
  & $- g_{ij} \bar q_{Li}  \gM \tau^I q_{Lj} \, \mathcal{W}_\mu^I$
  & $g=g^\dagger$ \\[1mm]
$\gmu$ & $(\text{Adj},1)_0$
  & $- \left( g_{ij}^q \bar q_{Li} \gM \frac{\la}{2} q_{Lj} 
  + g_{ij}^u \bar u_{Ri} \gM \frac{\la}{2} u_{Rj} 
  + g_{ij}^d \bar d_{Ri} \gM \frac{\la}{2} d_{Rj} \right) \mathcal{G}_\mu^a$ & $g=g^\dagger$ \\[1mm]
$\hmu$ & $(\text{Adj},\text{Adj})_0$
  & $- g_{ij} \bar q_{Li}  \gM \tau^I \frac{\la}{2} q_{Lj} \, \mathcal{H}_\mu^{aI}$ & $g=g^\dagger$ \\[1mm]
$\qmu$ & $(3,2)_{-\frac{5}{6}}$
  & $-g_{ij} \varepsilon_{abc} \bar u_{Rib} \gM \epsilon q_{Ljc}^c \, \mathcal{Q}_\mu^{5a\dagger} + \text{h.c.}$ & -- \\[1mm]
$\ymu$ & $(\bar 6,2)_{-\frac{5}{6}}$
  & $-g_{ij} \oh \left[ \bar u_{Ria} \gM \epsilon q_{Ljb}^c + 
  \bar u_{Rib} \gM \epsilon q_{Lja}^c \right] \mathcal{Y}_\mu^{5ab\dagger}  + \text{h.c.}$ & -- \\[1mm]
$\phi$ & $(1,2)_{-\frac{1}{2}}$
  & $- g_{ij}^u \bar q_{Li} u_{Rj} \, \phi - g_{ij}^d \bar q_{Li} d_{Rj} \, \tilde \phi  + \text{h.c.}$ & -- \\[1mm]
$\Phi$ & $(\text{Adj},2)_{-\frac{1}{2}}$
  & $- g_{ij}^u \bar q_{Li} \frac{\la}{2} u_{Rj} \, \Phi^a - g_{ij}^d \bar q_{Li} \frac{\la}{2} d_{Rj} \, \tilde \Phi^a  + \text{h.c.}$ & -- \\[1mm]
$\Of$ & $(\bar 6,1)_{-\frac{4}{3}}$
  & $-g_{ij} \oh \left[ \bar u_{Ria} u_{Rjb}^c + 
  \bar u_{Rib} u_{Rja}^c \right] \Omega^{4ab\dagger} + \text{h.c.}$ & $g=g^T$ \\[1mm]
$\So$ & $(\bar 6,\text{Adj})_{-\frac{1}{3}}$
  & $-g_{ij} \oh \left[ \bar q_{Lia} \tI \epsilon q_{Ljb}^c + 
  \bar q_{Lib} \tI \epsilon q_{Lja}^c \right] \So^{Iab\dagger} + \text{h.c.}$ & $g=g^T$
\end{tabular}
\end{center}
\caption{Vector bosons and scalar representations mediating $uu \to tt$.}
\label{tab:lagr}
\end{table}
We use standard notation with left-handed doublets $q_{Li}$ and right-handed singlets $u_{Ri}$, $d_{Ri}$; $\tI$ are the Pauli matrices, $\la$ the Gell-Mann matrices normalised to $\text{tr}(\la \lambda^b) = 2 \delta_{ab}$ and $\tilde \phi = \epsilon \phi$, $\psi^c = C \bar \psi^T$, with $\epsilon=i\tau^2$ and $C$ the charge conjugation matrix. The subindices $a,b,c$ denote colour. The bosons $\qmu$, $\ymu$, $\Of$, $\So$ are created by $uu$ fusion and exchanged in the $s$ channel, while the rest are exchanged in the $t$ (and $u$) channels.

If the new particles are heavy, their contribution to $uu \to tt$ can be described by an effective low-energy Lagrangian.
There are only five independent four-fermion operators contributing to $uu \to tt$~\cite{AguilarSaavedra:2010zi}, which can be taken as $O_{qq}^{1313}$, $O_{qq'}^{1313}$, $O_{uu}^{1313}$, $O_{qu}^{1313}$ and $O_{qu'}^{1313}$ (see appendix~\ref{sec:a}). The coefficients of these operators are given in Table~\ref{tab:Cttlike} for each of the vector boson and scalar representations in Table~\ref{tab:lagr}.\footnote{For $\wmu$ and $\hmu$ the normalisation in the Lagrangian differs from Ref.~\cite{delAguila:2010mx} by a factor of two, to simplify the presentation of the limits.}  The new physics scale $\Lambda$ equals the mass of the new boson or multiplet in each case.

\begin{table}[p]
\begin{center}
\begin{tabular}{c|ccccc}
& $C_{qq}^{1313}$ & $C_{qq'}^{1313}$ & $C_{uu}^{1313}$ & $C_{qu}^{1313}$ & $C_{qu'}^{1313}$ \\[1mm]
\hline \\[-4mm]
$\bmu$ & $-(g_{13}^q)^2$ & -- & $-(g_{13}^u)^2$ & -- & $2 g_{13}^q g_{13}^u$ \\[1mm]
$\wmu$ & $g_{13}^2$ & $- 2 g_{13}^2$ & -- & -- & -- \\[1mm]
$\gmu$ & $\os (g_{13}^q)^2$ & $-\oh (g_{13}^q)^2$ & $-\ot (g_{13}^u)^2$
  & $g_{13}^q g_{13}^u$ & $-\ot g_{13}^q g_{13}^u$  \\[1mm]
$\hmu$ & $-\ssx g_{13}^2$ & $\fsx g_{13}^2$ & -- & -- & -- \\[1mm]
$\qmu$ & -- & -- & -- & $2 g_{11} g_{33}^*$ & $-2 g_{11} g_{33}^*$  \\[1mm]
$\ymu$ & -- & -- & -- & $- g_{11} g_{33}^*$ & $- g_{11} g_{33}^*$ \\[1mm]
$\phi$ & -- & -- & -- & $g_{13}^u g_{31}^{u*}$ & -- \\[1mm]
$\Phi$ & -- & -- & -- & $-\os g_{13}^u g_{31}^{u*}$ & $\oh g_{13}^u g_{31}^{u*}$ \\[1mm]
$\Of$ & -- & -- & $g_{11} g_{33}^*$ & -- & -- \\[1mm]
$\So$ & $g_{11} g_{33}^*$ & $g_{11} g_{33}^*$ & -- & -- & --
\end{tabular}
\end{center}
\caption{Effective operator coefficients involved in like-sign $tt$ production for each vector boson and scalar representation. The new physics scale $\Lambda$ equals the mass of the new particle or multiplet.}
\label{tab:Cttlike}
\end{table}

Within this model-independent approach, the cross section for $uu \to tt$ can be compactly written in terms of five effective operator coefficients, the new physics scale $\Lambda$ and numerical constants $E_{1-3}$ that result from phase space integration and convolution with parton density functions (PDFs),
\begin{eqnarray}
\sigma(tt) & = & \frac{E_1}{\Lambda^4} \left[ |\Cqq^{1313}+\Cqqp^{1313}|^2 + |\Cuu^{1313}|^2 \right] \notag \\
& & + \frac{E_2}{\Lambda^4} \left[ |\Cqup^{1313}|^2 + |\Cqu^{1313}|^2 + \twt \, \RE \Cqup^{1313} \Cqu^{1313*} \right] \notag \\
& & + \frac{E_3}{\Lambda^4} \left\{ \RE \Cqup^{1313} \Cqu^{1313*}
+ \os \left[ |\Cqup^{1313}|^2 + |\Cqu^{1313}|^2 \right] \right\} \,.
\label{ec:xs}
\end{eqnarray}
The lowest order contributions arise at order $1/\Lambda^4$, since the SM amplitudes vanish~\cite{AguilarSaavedra:2010sq}. Clearly, higher-order operators can be neglected as long as the extra particles are heavy enough. In Appendix~B we discuss in more detail the range of validity of this approximation.
 
For Tevatron, we find\footnote{Equation~(\ref{ec:xs}) and these new values of $E_{1-3}$ correct the result previously given in Ref.~\cite{AguilarSaavedra:2010zi}, which had some missing contributions.} $E_1 = 14.5$, $E_2 = 2.13$, $E_3 = -1.95$ in units of $\text{fb} \cdot \text{TeV}^4$,
and the same factors for $\bar t \bar t$ production. We have taken $m_t = 172.5$ GeV and used CTEQ6L1 PDFs~\cite{Pumplin:2002vw} with $Q=m_t$. The CDF collaboration has also set individual limits on $t_L t_L+\bar t_L \bar t_L$, $t_R t_R+\bar t_R \bar t_R$ and $t_L t_R+\bar t_L \bar t_R$ production (the subindices refer to the top quark chiralities), which imply
\begin{equation}
|\Cqq^{1313}+\Cqqp^{1313}| \leq 4.1~\text{TeV}^{-2} ~,\quad 
|\Cuu^{1313}| \leq 3.7~\text{TeV}^{-2} ~,\quad
|C_{qu^{(')}}^{1313}| \leq 11.3~\text{TeV}^{-2} \,.
\label{ec:lim2}
\end{equation}
For LHC with 7 TeV, we have $E_1 = 16.0$, $E_2 = 2.06$, $E_3 = -0.416$ in units of $\text{pb} \cdot \text{TeV}^4$, while for $\bar t \bar t$ production the corresponding factors are a factor of 100 smaller. An expected limit on this process from 2010 data can be
estimated from the recent CMS search for fourth generation quarks in the like-sign dilepton and trilepton final states, using 35 pb$^{-1}$~\cite{Chatrchyan:2011em} (see also Ref.~\cite{Rajaraman:2011rw}). We conservatively assume one (background) observed event, and for the signal in the dilepton decay channel a detection efficiency of 80\% for each lepton and 50\% for event reconstruction, which yields an overal efficiency of 1.6\%, including the branching ratio for $W \to e\nu,\mu \nu$. Using Feldman-Cousins statistics~\cite{Feldman:1997qc}, this translates into the 95\% CL upper limit
\begin{equation}
\sigma(tt) < 7.5~\text{pb} \quad \quad (\text{LHC estimate}) \,.
\label{ec:limLHC}
\end{equation}

\begin{figure}[htb]
\begin{center}
\begin{tabular}{ccc}
\epsfig{file=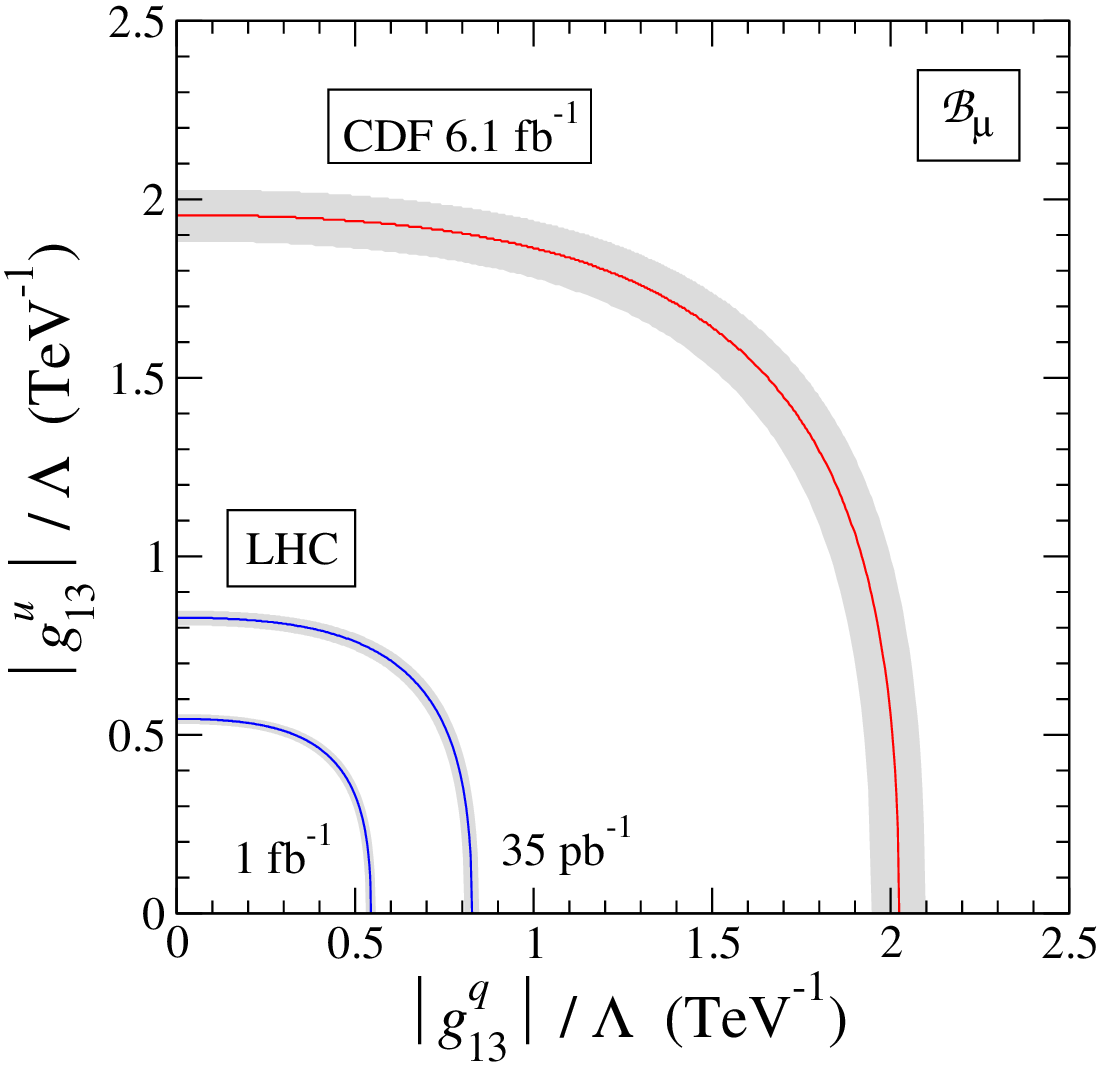,height=6cm,clip=} & \quad
\epsfig{file=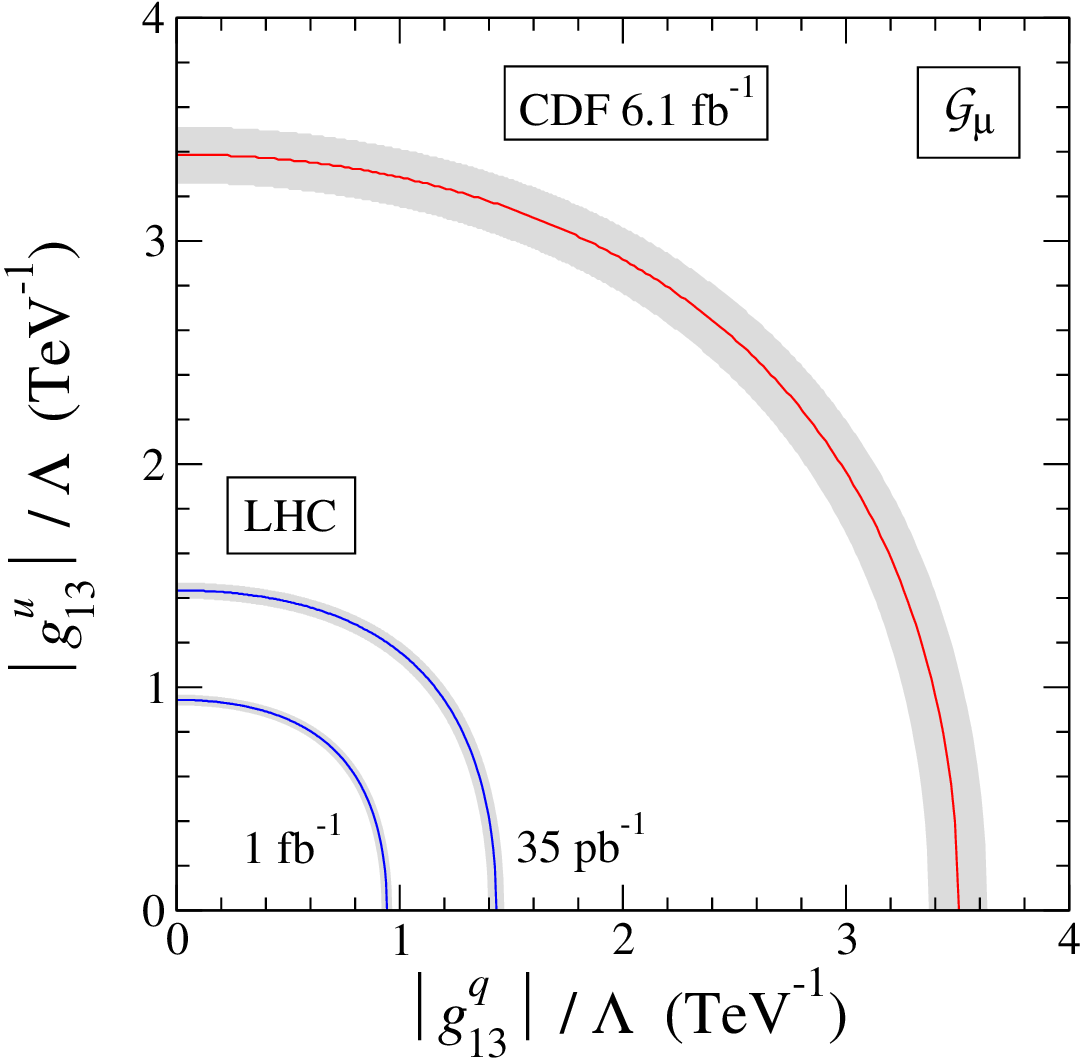,height=6cm,clip=}
\end{tabular}
\end{center}
\caption{Upper limits on the couplings for $\bmu$ and $\gmu$ vector boson representations.}
\label{fig:lim-BG}
\end{figure}

For heavy vector bosons $\bmu$ and $\gmu$ the reported CDF upper limits imply correlated constraints on the left-handed FCN couplings $g_{13}^q$ and the right-handed ones $g_{13}^u$ . The resulting two-dimensional upper bounds are shown in Fig.~\ref{fig:lim-BG}. In each plot, the red line corresponds to the central value directly obtained from Eq.~(\ref{ec:xs}), while the gray band represents the theoretical uncertainty obtained by varying the factorisation scale between $Q=2m_t$ and $Q=m_t/2$.
The outer blue line corresponds to the expected LHC limit with 35 pb$^{-1}$, and the inner one to an estimated limit $\sigma(tt) < 1.4~\text{pb}$ for 1 fb$^{-1}$, obtained with a naive luminosity rescaling of Eq.~(\ref{ec:limLHC}). The gray bands again correspond to the theoretical uncertainty in the cross section. We clearly observe that the analysis of existing 2010 LHC data could greatly improve present bounds.

For the remaining vector boson representations and for all scalars, the upper limits on $tt$ production translate into bounds on either a single FCN coupling ($\wmu$ and $\hmu$), a product of two FCN couplings ($\phi$ and $\Phi$), or a product of two flavour-diagonal couplings ($\qmu$, $\ymu$, $\Of$, $\So$). They are collected in Table~\ref{tab:lim-rest}, including the uncertainty corresponding to the variation of factorisation scale.

\begin{table}[htb]
\begin{center}
\begin{tabular}{c|ccccccc}
& & & & \multicolumn{2}{c}{LHC expected} \\[-1mm]
& & & CDF limit & 35 pb$^{-1}$ & 1 fb$^{-1}$ \\[1mm]
\hline \\[-4mm]
$\wmu$ & $|g_{13}|/\Lambda$ & $<$ & $2.02^{+0.07}_{-0.08}$
                            & $0.827^{+0.020}_{-0.021}$ & $0.544^{+0.013}_{-0.014}$
                            & TeV$^{-1}$ \\[1mm]
$\hmu$ & $|g_{13}|/\Lambda$ & $<$ & $3.50^{+0.13}_{-0.13}$ 
                            & $1.433^{+0.034}_{-0.037}$ & $0.942^{+0.022}_{-0.024}$
                            & TeV$^{-1}$ \\[1mm]
$\qmu$ & $|g_{11} g_{33}|/\Lambda^2$ & $<$ & $3.72^{+0.26}_{-0.27}$
                                & $0.716^{+0.038}_{-0.039}$ & $0.310^{+0.017}_{-0.017}$ 
                                & TeV$^{-2}$ \\[1mm]
$\ymu$ & $|g_{11} g_{33}|/\Lambda^2$ & $<$ & $8.6^{+0.7}_{-0.9}$
                                & $1.32^{+0.06}_{-0.06}$ & $0.568^{+0.025}_{-0.027}$
                                & TeV$^{-2}$ \\[1mm]
$\phi$ & $|g_{13}^u g_{31}^u|/\Lambda^2$ & $<$ & $11.2^{+0.8}_{-0.8}$
                                & $1.94^{+0.09}_{-0.10}$ & $0.838^{+0.040}_{-0.043}$
                                & TeV$^{-2}$ \\[1mm]
$\Phi$ & $|g_{13}^u g_{31}^u|/\Lambda^2$ & $<$ & $21.3^{+1.6}_{-1.6}$
                                & $3.67^{+0.18}_{-0.19}$ & $1.59^{+0.08}_{-0.08}$
                                & TeV$^{-2}$ \\[1mm]
$\Of$ & $|g_{11} g_{33}|/\Lambda^2$ & $<$ & $3.79^{+0.27}_{-0.28}$
                                & $0.684^{+0.033}_{-0.035}$ & $0.296^{+0.014}_{-0.015}$
                                & TeV$^{-2}$ \\[1mm]
$\So$ & $|g_{11} g_{33}|/\Lambda^2$ & $<$ & $2.04^{+0.15}_{-0.15}$
                                & $0.342^{+0.017}_{-0.017}$ & $0.148^{+0.007}_{-0.008}$
                                & TeV$^{-2}$
\end{tabular}
\end{center}
\caption{Upper limits on the couplings for the remaining vector boson and scalar representations.}
\label{tab:lim-rest}
\end{table}

\section{Implications on the $t \bar t$ asymmetry}
\label{sec:3}

In a general dimension-six effective Lagrangian, the gauge-invariant four-fermion operators contributing to $uu \to tt$ are independent from the ones involved in $u\bar u,d\bar d \to t \bar t$. Thus, a general model-independent connection between both processes only based on gauge symmetry does not exist. Nevertheless, when one considers extensions of the SM with explicit representations of vector bosons or scalars, a direct relation between the coefficients of these operators can be established in some cases.

The $t\bar t$ cross section and FB asymmetry including four-fermion operator contributions up to order $1/\Lambda^4$ have been given in Refs.~\cite{AguilarSaavedra:2010zi,AguilarSaavedra:2011vw}, and we omit them here for brevity (see also Ref.~\cite{Delaunay:2011gv},
and Refs.~\cite{Jung:2009pi} for $1/\Lambda^2$ calculations).
On the other hand, the explicit
operator coefficients corresponding to all possible vector boson and scalar representations have been obtained in Ref.~\cite{AguilarSaavedra:2011vw}. (For vector bosons they were previously given in Ref.~\cite{delAguila:2010mx}.) We collect them in Tables~\ref{tab:CAint}--\ref{tab:CA3}, but including only the representations relevant for like-sign top production.
A comparison with the coefficients of operators mediating $tt$ production in Table~\ref{tab:Cttlike} allows to find the relation between both processes, if any. 

\begin{table}[htb]
\begin{center}
\begin{small}
\begin{tabular}{c|ccccccc}
& $C_{qq}^{3113}$  & $C_{qq'}^{1133}$ & $C_{uu}^{3113}$
& $C_{ud'}^{3311}$ & $C_{qu}^{1331}$  & $C_{qu}^{3113}$ 
& $C_{qd}^{3113}$ \\[1mm]
\hline \\[-4mm]
$\bmu$ & $-|{\color{blue} g_{13}^q}|^2$ & -- & $-|{\color{blue} g_{13}^u}|^2$ & -- & -- & -- & -- \\[1mm]
$\wmu$ & $|{\color{blue} g_{13}}|^2$ & $-2 |{\color{blue} g_{13}}|^2$ & -- & -- & -- & -- & -- \\[1mm]
$\gmu$ & $\os |{\color{blue} g_{13}^q}|^2$ & $-\oh g_{11}^q g_{33}^q$ 
& \!\!\begin{tabular}{c}$\os |{\color{blue} g_{13}^u}|^2$ \\ $-\oh g_{11}^u g_{33}^u$
\end{tabular}\!\!
& $-\of g_{33}^u g_{11}^d$ & $\oh g_{11}^q g_{33}^u$ & $\oh g_{33}^q g_{11}^u$ & $\oh g_{33}^q g_{11}^d$ \\[1mm]
$\hmu$ 
& \!\!\begin{tabular}{c}$-\os |{\color{blue} g_{13}}|^2$ \\ $- g_{11} g_{33}$
\end{tabular}\!\!
& \!\!\begin{tabular}{c}$\ot |{\color{blue} g_{13}}|^2$ \\ $+\oh g_{11} g_{33}$
\end{tabular}\!\!
& -- & -- & -- & -- & --\\[1mm]
$\qmu$ & -- & -- & -- & -- & $|g_{31}|^2$ & $|g_{13}|^2$ & -- \\[1mm]
$\ymu$ & -- & -- & -- & -- & $-\oh|g_{31}|^2$ & $-\oh|g_{13}|^2$\\[1mm]
$\phi$ & -- & -- & -- & -- & $\oh |{\color{blue} g_{13}^u}|^2$ & $\oh |{\color{blue} g_{31}^u}|^2$ & $\oh |g_{31}^d|^2$\\[1mm]
$\Phi$ & -- & -- & -- & -- & $-\otw |{\color{blue} g_{13}^u}|^2$ & $-\otw |{\color{blue} g_{31}^u}|^2$ & $-\otw |g_{31}^d|^2$\\[1mm]
$\Of$ & -- & -- & $|g_{13}|^2$ & -- & -- & -- & -- \\[1mm]
$\So$ & $|g_{13}|^2$ & $|g_{13}|^2$ & -- & -- & -- & -- & --
\end{tabular}
\end{small}
\end{center}
\caption{Coefficients of effective operators interfering with the SM amplitudes for $u \bar u,d\bar d \to t \bar t$. The new physics scale $\Lambda$ equals the mass of the new particle or multiplet.}
\label{tab:CAint}
\end{table}

\begin{table}[p]
\begin{center}
\begin{tabular}{c|ccccccc}
& $C_{qq}^{1133}$ & $C_{qq'}^{3113}$ & $C_{uu}^{1133}$
& $C_{ud}^{3311}$ \\[1mm]
\hline \\[-4mm]
$\bmu$ & $-g_{11}^q g_{33}^q$ & -- & $-g_{11}^u g_{33}^u$ & $-\oh g_{33}^u g_{11}^d$ \\[1mm]
$\wmu$ & $g_{11}^q g_{33}^q$ & $-2 g_{11}^q g_{33}^q$ & -- & -- \\[1mm]
$\gmu$ & $\os g_{11}^q g_{33}^q$ & $-\oh |{\color{blue} g_{13}^q}|^2$ 
& \!\!\begin{tabular}{c}$\os g_{11}^u g_{33}^u$ \\ $-\oh |{\color{blue} g_{13}^u}|^2$
\end{tabular}
& $\otw g_{33}^u g_{11}^d$ \\[1mm]
$\hmu$
& \!\!\begin{tabular}{c}$-\os g_{11} g_{33}$ \\ $- |{\color{blue} g_{13}}|^2$
\end{tabular}
& \!\!\begin{tabular}{c}$\oh |{\color{blue} g_{13}}|^2$ \\ $+\ot g_{11} g_{33}$
\end{tabular}
& -- & -- \\[1mm]
%
%
%
%
%
$\Of$ & -- & -- & $|g_{13}|^2$ & -- \\[1mm]
$\So$ & $|g_{13}|^2$ & $|g_{13}|^2$ & -- & --
\end{tabular}
\end{center}
\caption{Coefficients of $\bar L L \bar L L$ and $\bar R R \bar R R$ effective operators contributing to $u \bar u,d\bar d \to t \bar t$ at quadratic level. For the representations not listed all these coefficients vanish. The new physics scale $\Lambda$ equals the mass of the new particle or multiplet.}
\label{tab:CA1}
\end{table}

\begin{table}[p]
\begin{center}
\begin{tabular}{c|ccccccc}
& $C_{qu}^{3311}$ & $C_{qu'}^{1331}$ & $C_{qu'}^{3113}$ & $C_{qu'}^{3311}$
& $C_{qd'}^{3113}$ \\[1mm]
\hline \\[-4mm]
$\bmu$ & -- & $g_{11}^q g_{33}^u$ & $g_{33}^q g_{11}^u$ & $2 {\color{blue} g_{13}^{q*} g_{13}^u}$ & $g_{33}^q g_{11}^d$ \\[1mm]
%
%
$\gmu$ & ${\color{blue} g_{13}^{q*} g_{13}^u}$ & $-\os g_{11}^q g_{33}^u$ & $-\os g_{33}^q g_{11}^u$ & $-\ot {\color{blue} g_{13}^{q*} g_{13}^u}$ & $-\os g_{33}^q g_{11}^d$ \\[1mm]
%
%
$\qmu$ & $2 g_{13} g_{31}^*$ & $-|g_{31}|^2$ & $-|g_{13}|^2$ &
$-2 g_{13} g_{31}^*$ & -- \\[1mm]
$\ymu$  & $- g_{13} g_{31}^*$ & $-\oh |g_{31}|^2$ & $-\oh |g_{13}|^2$ & 
$-g_{13} g_{31}^*$ & -- \\[1mm]
$\phi$ & $g_{11}^{u*} g_{33}^u$ & -- & -- & -- & -- \\[1mm]
$\Phi$ & $-\os g_{11}^{u*} g_{33}^u$ & $\of |{\color{blue} g_{13}^u}|^2$ & $\of |{\color{blue} g_{31}^u}|^2$ & $\oh g_{11}^{u*} g_{33}^u$ & $\of |g_{31}^d|^2$ \\[1mm]
%
%
\end{tabular}
\end{center}
\caption{Coefficients of $\bar L R \bar R L$ effective operators contributing to $u \bar u,d\bar d \to t \bar t$ at quadratic level. For the representations not listed all these coefficients vanish. The new physics scale $\Lambda$ equals the mass of the new particle or multiplet.}
\label{tab:CA2}
\end{table}

\begin{table}[p]
\begin{center}
\begin{tabular}{c|ccccccc}
& $C_{qq\epsilon}^{1331}$ & $C_{qq\epsilon}^{3311}$ 
& $C_{qq\epsilon'}^{1331}$ & $C_{qq\epsilon'}^{3311}$ \\
\hline \\[-4mm]
%
%
%
%
%
%
$\phi$ & ${\color{blue} g_{13}^u} g_{31}^d$ & $g_{33}^u g_{11}^d$ & -- & -- \\[1mm]
$\Phi$ & $-\os {\color{blue} g_{13}^u} g_{31}^d$ & $-\os g_{33}^u g_{11}^d$ & $\oh {\color{blue} g_{13}^u} g_{31}^d$ & $\oh g_{33}^u g_{11}^d$ \\[1mm]
%
%
\end{tabular}
\end{center}
\caption{Coefficients of $\bar L R \bar L R$ effective operators contributing to $u \bar u,d\bar d \to t \bar t$ at quadratic level. For the representations not listed all these coefficients vanish. The new physics scale $\Lambda$ equals the mass of the new particle or multiplet.}
\label{tab:CA3}
\end{table}

For the singlets $\bmu$ ($Z^\prime$ bosons), the operators $O_{qq}^{3113}$, $O_{uu}^{3113}$ contributing at first order to $t \bar t$ production depend on the couplings $g_{13}^q$ and $g_{13}^u$ (see Table \ref{tab:CAint}), which are the very same couplings appearing in like-sign $tt$. (These and other couplings which are involved in $tt$ production are displayed in blue for a better identification.) This relation comes from the fact that these operators correspond to $t$-channel exchange of the vector boson in $u \bar u \to t \bar t$, which is related to $tt$ production by conjugation of the $\bar u \gM t \, \bmu$ vertex ($\bmu$ is real).
At the quadratic level we find, for example, the operators $O_{qq}^{1133}$, $O_{uu}^{1133}$, which involve diagonal couplings $g_{11}$, $g_{33}$. They correspond to $s$-channel exchange of the new boson and have no counterpart in $tt$ production.
Therefore, a direct relation among $t\bar t$ and $tt$ production exists when $s$-channel contributions to the former are absent or negligible.\footnote{It is possible to get around this relation and evade the constraints from $tt$ production by introducing two bosons in the same irreducible representation with degenerate masses and couplings that differ by a factor of $i$. Then, the respective contributions to $tt$ production cancel each other. This mechanism can be natural if both bosons belong to a multiplet of an extended (flavour) symmetry of the complete theory, as in the model in Ref.~\cite{Jung:2011zv}.}
This is indeed the case in several models proposed to accommodate the $t \bar t$ asymmetry
by the exchange of a $t$-channel $Z'$ boson. (The $s$-channel exchange of a $Z'$ could easily display a peak in the $t \bar t$ invariant mass, which is not observed in the measurements.)
For the colour octet $\gmu$, the situation is similar, except for the fact that in this case $s$-channel contributions interfere with the SM amplitude and appear already at first order, for example in the operators $O_{qq'}^{1133}$, $O_{ud'}^{3311}$. Again, the relation exists for $t$-channel exchange only.
The cases of the $\text{SU}(2)_L$ triplets $\wmu$ and $\hmu$ are analogous to the ones of the singlets $\bmu$ and $\gmu$, respectively but, in addition, a $t$-channel exchange of the charged member(s) of the multiplet in $d \bar d \to t \bar t$ is possible. This exchange is parameterised by the same operators as in $u \bar u \to t \bar t$, which in this case also contribute to the $t \bar t$ asymmetry through $d \bar d \to t \bar t$.

\begin{figure}[htb]
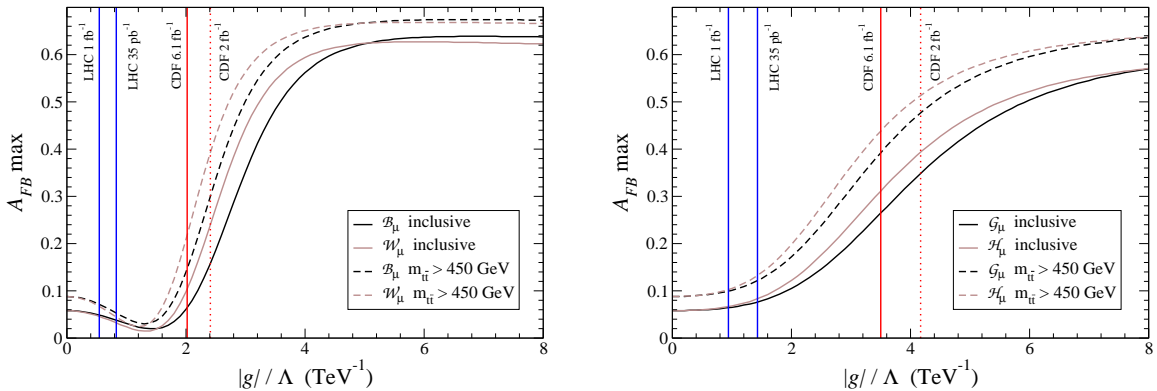

\begin{center}
\begin{tabular}{ccc}
\epsfig{file=Figs/AFB-Z,height=5.1cm,clip=} & \quad
\epsfig{file=Figs/AFB-G.eps,height=5.1cm,clip=}
\end{tabular}
\end{center}
\caption{Maximum FB asymmetry from $t$-channel vector boson exchange as a function of the limit on $|g|/\Lambda$. The vertical red line corresponds to the present upper limit from CDF, and the blue line to the expected LHC limit with 35 pb$^{-1}$.}
\label{fig:AFB-BG}
\end{figure}

In order to explicitly relate the new contributions to the asymmetry $\afb$ from $t$-channel exchange with the $tt$ cross section, we first make the following observations. For $\bmu$ and $\gmu$, the asymmetry depends on two couplings $g_{13}^q$, $g_{13}^u$ but its maximum value will be achieved when one of them is zero, once the like-sign $tt$ constraints are imposed, see Fig.~\ref{fig:lim-BG}. The reason is that the terms containing the product are forward-backward symmetric and dilute the asymmetry. We consider this most favourable scenario, and call $g$ the nonvanishing coupling. 
For $\wmu$ and $\hmu$, there is only one coupling $g=g_{13}$, which corresponds to the case $g_{13}^u = 0$ in the two-dimensional case.
With these clarifications in mind, we plot in Fig.~\ref{fig:AFB-BG} the maximum value of $A_\text{FB}$ achievable as a function of the limit on $|g|/\Lambda$.
The solid red lines are the limits implied by the CDF measurement, while the red dotted lines correspond to a previous one $\sigma(ttX + \bar t \bar t X) < 980~\text{fb}^{-1}$~\cite{Aaltonen:2008hx} from a search for flavour-violating scalars $\phi$ with 2 fb$^{-1}$. The blue lines on the left are the expected bounds from LHC. From the left plot, it is apparent that $t$-channel exchange of a heavy $Z'$ cannot accommodate the large asymmetry $A_\text{FB} = 0.475 \pm 0.114$ for $m_{t\bar t} > 450$ GeV~\cite{Aaltonen:2011kc} measured at Tevatron. (We can also observe that this strong conclusion could not be drawn from the previous 2 fb$^{-1}$ limit, in spite of recent claims~\cite{Shu:2011au}.)
We should, on the other hand, remember that for lighter $Z'$ bosons the effective operator approximation overestimates the $tt$ cross section. We present in appendix~\ref{sec:b} a comparison with exact results showing that masses $M_{Z'} \sim 200$ GeV are in the borderline of exclusion by CDF data, and lighter $Z'$ bosons are still (barely) allowed. With the more stringent LHC limits they will be definitely seen or excluded. This last remark applies to the $t$-channel exchange of colour octets, too.

\begin{figure}[htb]
\begin{center}
\epsfig{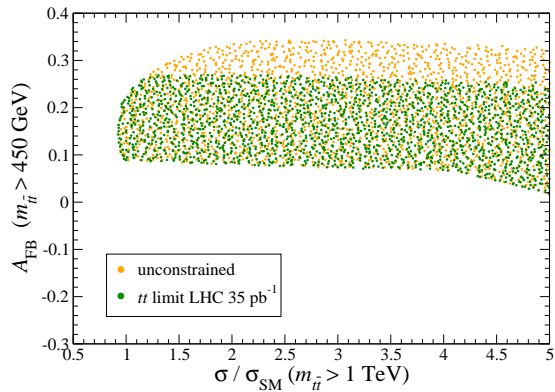}
\end{center}
\caption{Allowed regions for the Tevatron $t \bar t$ asymmetry and the $t \bar t$ tail at LHC for a colour-singlet scalar $\phi$.}
\label{fig:phi}
\end{figure}

The coefficients for $t \bar t$ and $tt$ production are not related for the colour-triplet and sextet vector bosons, $\qmu$ and $\ymu$, nor for any of the scalars. In the case of the vectors $\qmu$ and $\ymu$ and the colour-sextet scalars $\Of$ and $\So$, the reason is simply that these new particles contribute in $u$ channel to $t \bar t$ production, involving off-diagonal couplings $g_{13}$, and in $s$ channel to $tt$, with diagonal couplings $g_{11}$, $g_{33}$. (Realistic models will generally have both but diagonal ones may be suppressed~\cite{Grinstein:2011yv}.) 
On the other hand, the absence of a direct relation for the isodoublets $\phi$ and $\Phi$---which might have been expected in view of the results for $\bmu$ and $\gmu$---is explained by the fact that these scalars are in complex representations of the SM gauge group.
Still, the negative limits on $tt$ production constrain the couplings of these scalars. For the colour singlet $\phi$, this has interesting implications for the FB asymmetry. We have recently shown in Ref.~\cite{AguilarSaavedra:2011vw} that $\phi$ allows to reproduce an asymmetry $\afb \sim 0.3$ with a moderate $t \bar t$ tail at LHC, $\sigma \simeq 1.5\, \sigma_\text{SM}$ for $m_{t \bar t} > 1$ TeV. This can be achieved with $g_{13}^u/\Lambda$, $g_{31}^u/\Lambda$ of the same order and not much larger than $1~\text{TeV}^{-1}$ (otherwise the quadratic terms dilute the asymmetry). However, this possibility may be precluded by $tt$ production, see Table~\ref{tab:lim-rest}. The effect of a limit on $tt$ production can be clearly seen in Fig.~\ref{fig:phi}, where we show the relation between $\afb$ and the $t \bar t$ tail at LHC in two scenarios: letting all the six couplings arbitrary and imposing the potential LHC bound for 35 pb$^{-1}$ in Table~\ref{tab:lim-rest}. We can thus conclude that, if $tt$ production is not observed in 2010 LHC data, an extra (heavy) scalar doublet is not sufficient to explain the present $\afb$ measured at Tevatron.

\section{Discussion}
\label{sec:4}

Using the upper limit on $tt$ production at Tevatron reported by the CDF collaboration, Eq.~(\ref{ec:lim}), we have set bounds on the masses and couplings of all the extra vector and scalar bosons that can mediate the process $uu \to tt$. This includes popular models, such as SM extensions with a new $Z'$~\cite{Langacker:2000ju} or a general two-Higgs doublet model~\cite{Atwood:1996vj}. The flavour constraints obtained here from top physics are complementary to the ones from low-energy observables. Moreover, there are other exotic, loosely bound models to which these limits also apply~\cite{Berger:2010fy}.

The limits presented here have been obtained using an effective operator framework, assuming that the masses $M$ of the new states are large, as compared to the momentum transfer. The range of validity of the approximation is studied in Appendix~\ref{sec:b}. In the case of new particles exchanged in the $s$ channel, the limits so derived are conservative. In particular, if the new states are light enough to be directly produced on-shell, the suppression will not be $1/M^4$ as in the off-shell case but $1/M\Gamma$, where $\Gamma$ is the width of the $s$-channel resonance. Therefore, a dedicated search covering all possible vector boson ($\qmu$, $\ymu$) and scalar ($\Of$, $\So$) representations will be welcome to get more stringent limits for small couplings and masses. On the other hand, the actual limits on {\em light} extra particles exchanged in the $t$ channel will be weaker than the ones we give. For this reason, we provide exact results for a light $Z'$ in Appendix~\ref{sec:b}.

A second, interesting consequence of the CDF limit concerns the FB asymmetry in $t \bar t$ production. The present discrepancy between its measured value~\cite{Aaltonen:2011kc} and the SM expectation has motivated a profusion of models that, while keeping the $t \bar t$ cross section close to the SM value (consistent with the measurements), accommodate a much larger asymmetry. Like-sign top pair production can be used to constrain the parameter space of some of these models. In particular, models that enhance the asymmetry by the $t$-channel exchange of a single $Z'$ boson~\cite{Jung:2009jz} give a correlated rate of like-sign top pair production. This relation, with the prediction of a striking effect to be seen at LHC, has been pointed out before~\cite{Jung:2009jz,Cao:2009uz,Choudhury:2010cd,
Cao:2011ew,Bhattacherjee:2011nr,Berger:2011ua}.\footnote{In the revised version of Ref.~\cite{Berger:2011ua}, which appeared on the arXiv the same day as the first version of the present paper, the implications of the first CDF limit~\cite{Aaltonen:2008hx} have also been included.}
But, remarkably, the absence of like-sign $tt$ production at Tevatron already rules out these models as the only explanation of the $t \bar t$ asymmetry, except for very light $Z'$ bosons. Here it is worth mentioning that a recent CDF dijet excess~\cite{Aaltonen:2011mk} could be interpreted as a hint of a new light $Z'$ boson, see for example Ref.~\cite{Buckley:2011vc}. In this paper we have seen that a $Z'$ with this mass might partially account for the FB asymmetry and still give $tt$ pairs with a rate below the present CDF bound.

The Tevatron limit on like-sign $tt$ production can certainly be improved at LHC, even with small statistics, due to the fact that the latter is a $pp$ collider. We have estimated that by analysing 2010 data, a limit of $\sigma(tt) < 7.5$ pb might be achieved. In case that a $tt$ excess is not observed, this would rule out all models with a single ---heavy or light--- $Z'$  as candidates to explain the asymmetry. (As we have already mentioned, more complex models with extra symmetries~\cite{Jung:2011zv} can evade this conclusion thanks to a cancellation of the different contributions.) In addition, it would exclude sizeable $t$-channel contributions to the asymmetry from scalar doublets $\phi$ and colour octet bosons, and thus further constrain the parameter space of these models.

\section*{Acknowledgements}

We thank D. Whiteson for useful discussions.
This work has been partially supported by projects FPA2010-17915 (MICINN), FQM 101 and FQM 437 (Junta de Andaluc\'{\i}a) and CERN/FP/116397/2010 (FCT).

\appendix
\section{Four-fermion operators for $tt$ and $t \bar t$ production.}
\label{sec:a}

We use the minimal basis in Ref.~\cite{AguilarSaavedra:2010zi} for gauge-invariant four-fermion operators. Fermion fields are ordered according to their spinorial index contraction, and
subindices $a$, $b$ indicate the pairs with colour indices contracted, if this pairing is different from the one for the spinorial contraction. Our basis consists of the following operators:

\noindent
(i) $\bar L L \bar L L$ operators
\begin{align}
& O_{qq}^{ijkl} = \oh (\bar q_{Li} \gM q_{Lj}) (\bar q_{Lk} \gm q_{Ll}) \,,
&& O_{qq'}^{ijkl} = \oh (\bar q_{Lia} \gM q_{Ljb}) (\bar q_{Lkb} \gm q_{Lla}) \,,
\notag \\
& O_{\ell q}^{ijkl} = (\bar \ell_{Li} \gM \ell_{Lj}) (\bar q_{Lk} \gm q_{Ll}) \,,
&& O_{\ell q'}^{ijkl} = (\bar \ell_{Li} \gM q_{Lj}) (\bar q_{Lk} \gm \ell_{Ll}) \,,
\notag \\
& O_{\ell \ell}^{ijkl} = \oh (\bar \ell_{Li} \gM \ell_{Lj}) (\bar \ell_{Lk} \gm \ell_{Ll}) \,.
\label{ec:LLLL}
\end{align}
(ii) $\bar R R \bar R R$ operators
\begin{align}
& O_{uu}^{ijkl} = \oh (\bar u_{Ri} \gM u_{Rj}) (\bar u_{Rk} \gm u_{Rl}) \,,
&& O_{dd}^{ijkl} = \oh (\bar d_{Ri} \gM d_{Rj}) (\bar d_{Rk} \gm d_{Rl}) \,,
\notag \\
& O_{ud}^{ijkl} = (\bar u_{Ri} \gM u_{Rj}) (\bar d_{Rk} \gm d_{Rl}) \,,
&& O_{ud'}^{ijkl} = (\bar u_{Ria} \gM u_{Rjb}) (\bar d_{Rkb} \gm d_{Rla}) \,,
\notag \\
& O_{eu}^{ijkl} = (\bar e_{Ri} \gM e_{Rj}) (\bar u_{Rk} \gm u_{Rl}) \,, 
&& O_{ed}^{ijkl} = (\bar e_{Ri} \gM e_{Rj}) (\bar d_{Rk} \gm d_{Rl}) \,,
\notag \\
& O_{ee}^{ijkl} = \oh (\bar e_{Ri} \gM e_{Rj}) (\bar e_{Rk} \gm e_{Rl}) \,.
\label{ec:RRRR}
\end{align}
(iii) $\bar L R \bar R L$ operators
\begin{align}
& O_{qu}^{ijkl} = (\bar q_{Li} u_{Rj}) (\bar u_{Rk} q_{Ll}) \,,
&& O_{qu'}^{ijkl} = (\bar q_{Lia} u_{Rjb}) (\bar u_{Rkb} q_{Lla}) \,,
\notag \\
& O_{qd}^{ijkl} = (\bar q_{Li} d_{Rj}) (\bar d_{Rk} q_{Ll}) \,,
&& O_{qd'}^{ijkl} = (\bar q_{Lia} d_{Rjb}) (\bar d_{Rkb} q_{Lla}) \,,
\notag \\
& O_{\ell u}^{ijkl} = (\bar \ell_{Li} u_{Rj}) (\bar u_{Rk} \ell_{Ll}) \,,
&& O_{\ell d}^{ijkl} = (\bar \ell_{Li} d_{Rj}) (\bar d_{Rk} \ell_{Ll}) \,,
\notag \\
& O_{qe}^{ijkl} = (\bar q_{Li} e_{Rj}) (\bar e_{Rk} q_{Ll}) \,,
&& O_{qde}^{ijkl} = (\bar \ell_{Li} e_{Rj}) (\bar d_{Rk} q_{Ll}) \,,
\notag \\
& O_{\ell e}^{ijkl} = (\bar \ell_{Li} e_{Rj}) (\bar e_{Rk} \ell_{Ll}) \,.
\label{ec:LRRL}
\end{align}
(iv) $\bar L R \bar L R$ operators
\begin{align}
& O_{qq\epsilon}^{ijkl} = (\bar q_{Li} u_{Rj}) \left[ (\bar q_{Lk} \epsilon)^T d_{Rl} \right] \,,
&& O_{qq\epsilon'}^{ijkl} = (\bar q_{Lia} u_{Rjb}) \left[ (\bar q_{Lkb} \epsilon)^T d_{Rla} \right] \,,
\notag \\
& O_{\ell q\epsilon}^{ijkl} = (\bar \ell_{Li} e_{Rj}) \left[ (\bar q_{Lk} \epsilon)^T u_{Rl} \right] \,,
&& O_{q \ell \epsilon}^{ijkl} = (\bar q_{Li} e_{Rj}) \left[ (\bar \ell_{Lk} \epsilon)^T u_{Rl} \right] \,.
\label{ec:LRLR}
\end{align}

\section{Exact calculations with flavour-violating $Z^\prime$}
\label{sec:b}

Several models reproduce the top FB asymmetry measured at Tevatron by introducing flavour-violating $Z'$ bosons with masses of just a few hundred GeV. In this case the approximation with dimension-six operators is not accurate. In this appendix, we perform the exact calculation for a $t$-channel $Z'$ of mass $M\equiv \Lambda$ in the representation $\bmu$, with only right-handed couplings.\footnote{Left-handed couplings are constrained by $B$ physics, so models in the literature usually assume right-handed couplings for $Z'$. For left-handed couplings the limits from like-sign $tt$ are slightly less restrictive, see Eqs.~(\ref{ec:lim2}).} The result is interesting on its own, and also provides a test of the range of validity of our general results for $t$-channel exchange in the effective formalism. Note that the operator approximation just amounts to substituting $1/(\Lambda^2-t)$ by $1/\Lambda^2$ in the propagator of the new particle, which appears squared in the $uu\to tt$ cross section. Our cross sections agree with Refs.~\cite{Bhattacherjee:2011nr,Berger:2011ua}. For example, for $M_{Z'}=1$ TeV and
$g_{13}^u=3$ we obtain for LHC $\sigma(tt) \times \text{Br}(W \to e \nu,\mu \nu)^2 = 24.8$ pb, in good agreement with Fig. 3 of Ref.~\cite{Bhattacherjee:2011nr}, and taking $g_{13}^u=g_W$ (being $g_W$ the weak coupling) we obtain $\sigma(tt) \times \text{Br}(W \to \mu \nu)^2 = 13.2$ fb, in agreement with Table I of Ref.~\cite{Berger:2011ua}. 

We plot the cross sections calculated using the $Z'$ boson and the corresponding four-fermion operator, as a function of the boson mass but keeping $C/\Lambda^2$ constant, where $C=C_{uu}^{3113}$ is the coefficient of the relevant operator. The operator predictions then remain flat, while the exact ones deviate from this limit for lower $\Lambda$. We take $C/\Lambda^2 = -5.8\, \mathrm{TeV}^{-2}$, which gives $\afb = 0.3$ in the effective operator approximation.

\begin{figure}[htb]
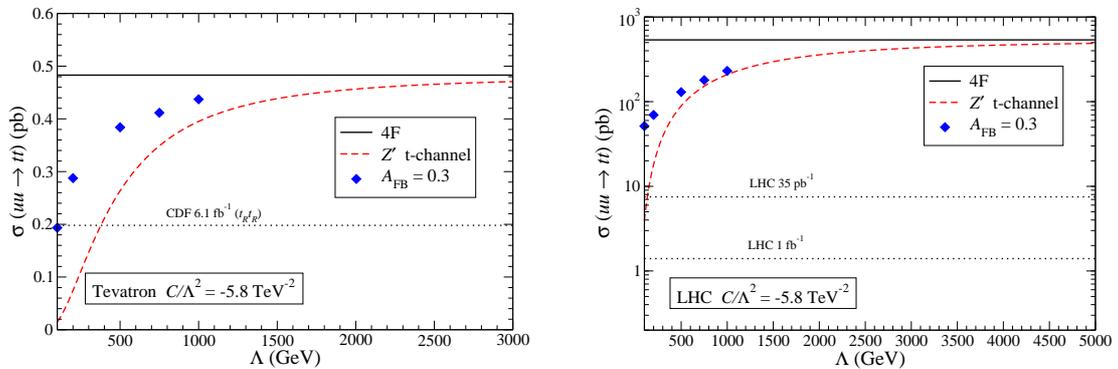

\begin{center}
\begin{tabular}{ccc}
\epsfig{file=Figs/xsec2-M,height=4.8cm,clip=} & \quad &
\epsfig{file=Figs/xsec7-M,height=4.8cm,clip=}
\end{tabular}
\caption{Comparison between four-fermion operator and exact calculations for a $t$ channel $Z'$ boson.}
\label{fig:comp}
\end{center}
\end{figure}
We see in Fig.\ \ref{fig:comp} that the exact calculation yields a smaller $tt$ cross section at Tevatron (left) and especially at LHC (right). As we have shown in Ref.~\cite{AguilarSaavedra:2011vw}, for small $\Lambda$ the assumed coupling $C/\Lambda^2=-5.8~\text{TeV}^{-2}$ is not sufficient to generate the required asymmetry;  
for this reason we have also calculated the values of $\sigma/\sigma_\text{SM}$ for $Z'$ masses $M=100,200,500,750,1000$ GeV and larger $|C|$ couplings, so as to reproduce $\afb = 0.3$. These points are displayed in both plots, together with the CDF ($t_R t_R$) and expected LHC limits. We observe that $Z'$ bosons with masses higher than $\sim 100\,\mathrm{GeV}$ and giving $\afb = 0.3$ generate too large a rate of $tt$ pairs, and would be excluded by the new CDF limits, were it not for a slightly reduced efficiency, discussed below. Moreover, for light $Z'$ bosons the contribution from on-shell production
$ug \to tZ' \to tt\bar u$ may be comparable and give an additional source of like-sign top pairs, strengthening the limits on the $Z'$ couplings. In this case the previous CDF limit on like-sign $tt$~\cite{Aaltonen:2008hx} already sets constraints on the model~\cite{Jung:2009jz}.
The right plot shows that even $Z'$ bosons of $100$ GeV can be easily probed at LHC with just 35~pb$^{-1}$. 

\begin{figure}[htb]
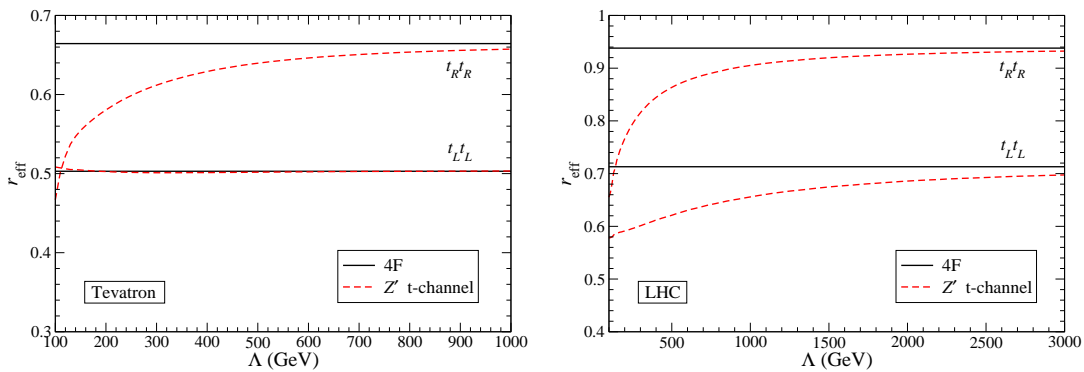

\begin{center}
\begin{tabular}{ccc}
\epsfig{file=Figs/eff-tev-M,height=4.8cm,clip=} &
\epsfig{file=Figs/eff-LHC-M,height=4.8cm,clip=}
\end{tabular}
\caption{Comparison between the charged lepton acceptance for four-fermion operators and a $t$-channel $Z'$ at Tevatron and LHC.}
\label{fig:comp2}
\end{center}
\end{figure}

We have also checked the possible efficiency variations when considering the four-fermion operators or the exact calculation, by considering the ratio
$r_\text{eff} = \sigma'/\sigma^\text{dil}$,
where $\sigma^\text{dil}$ is the total cross section times dileptonic branching ratio and $\sigma'$ the same cross section but also requiring for the charged lepton a pseudo-rapidity $|\eta| < 1.1$ ($|\eta| < 2.5$) for Tevatron (LHC) and transverse momentum $p_T > 20$ GeV in both cases. For completeness we investigate the efficiency variations for $t_L t_L$ production as well. For lower $\Lambda$ there is some efficiency decrease, see Fig.~\ref{fig:comp2}, because $tt$ pairs are more collinear due to the $t$-channel propagator enhancement at small production angles. The differences are larger for $t_R t_R$ production (which corresponds to the $Z'$ models with right-handed couplings usually considered) because charged leptons are emitted preferrably in the top flight direction, whereas for $t_L t_L$ production the charged leptons are emitted in the opposite one.

\begin{thebibliography}{99}


\bibitem{Berger:2010fy}
  E.~L.~Berger, Q.~H.~Cao, C.~R.~Chen, G.~Shaughnessy and H.~Zhang,
  Phys.\ Rev.\ Lett.\  {\bf 105} (2010) 181802
  {\tt [1005.2622 [hep-ph]]};
  H.~Zhang, E.~L.~Berger, Q.~H.~Cao, C.~R.~Chen and G.~Shaughnessy,
  Phys.\ Lett.\  B {\bf 696} (2011) 68
  {\tt [1009.5379 [hep-ph]]}.

\bibitem{Bauer:2009cc}
  C.~W.~Bauer, Z.~Ligeti, M.~Schmaltz, J.~Thaler and D.~G.~E.~Walker,
  Phys.\ Lett.\  B {\bf 690} (2010) 280
  {\tt [0909.5213 [hep-ph]]}.

\bibitem{CDFtt}
T. Aaltonen et al. [CDF Collaboration], CDF note 10466


\bibitem{delAguila:2010mx}
  F.~del Aguila, J.~de Blas and M.~P\'erez-Victoria,
  JHEP {\bf 1009} (2010) 033
  {\tt [1005.3998 [hep-ph]]}.

\bibitem{Aaltonen:2011kc}
  T.~Aaltonen {\it et al.}  [CDF Collaboration],
  {\tt 1101.0034 [hep-ex]}.


\bibitem{Jung:2009jz}
  S.~Jung, H.~Murayama, A.~Pierce and J.~D.~Wells,
  Phys.\ Rev.\  D {\bf 81} (2010) 015004
  {\tt [0907.4112 [hep-ph]]}.

\bibitem{Cao:2009uz}
  J.~Cao, Z.~Heng, L.~Wu and J.~M.~Yang,
  Phys.\ Rev.\  D {\bf 81}, 014016 (2010)
  {\tt [0912.1447 [hep-ph]]}.

\bibitem{Choudhury:2010cd}
  D.~Choudhury, R.~M.~Godbole, S.~D.~Rindani and P.~Saha,
  {\tt 1012.4750 [hep-ph]}.

\bibitem{Cao:2011ew}
  J.~Cao, L.~Wang, L.~Wu and J.~M.~Yang,
  {\tt 1101.4456 [hep-ph]}.

\bibitem{Bhattacherjee:2011nr}
  B.~Bhattacherjee, S.~S.~Biswal and D.~Ghosh,
  {\tt 1102.0545 [hep-ph]}.

\bibitem{Berger:2011ua}
  E.~L.~Berger, Q.~H.~Cao, C.~R.~Chen, C.~S.~Li and H.~Zhang,
  {\tt 1101.5625 [hep-ph]}.


\bibitem{Aaltonen:2011mk}
   T.~Aaltonen  [CDF Collaboration],
  {\tt 1104.0699 [hep-ex]}.


\bibitem{Jung:2011zv}
  S.~Jung, A.~Pierce and J.~D.~Wells,
  {\tt 1103.4835 [hep-ph]};




\bibitem{AguilarSaavedra:2011vw}
  J.~A.~Aguilar-Saavedra and M.~P\'erez-Victoria,
  {\tt 1103.2765 [hep-ph]}.

\bibitem{AguilarSaavedra:2010zi}
  J.~A.~Aguilar-Saavedra,
  Nucl.\ Phys.\  B {\bf 843} (2011) 638
  {\tt [1008.3562 [hep-ph]]}.

\bibitem{AguilarSaavedra:2010sq}
  J.~A.~Aguilar-Saavedra,
  {\tt 1008.3225 [hep-ph]}.

\bibitem{Pumplin:2002vw}
  J.~Pumplin, D.~R.~Stump, J.~Huston, H.~L.~Lai, P.~Nadolsky and W.~K.~Tung,
  JHEP {\bf 0207} (2002) 012
  {\tt [hep-ph/0201195]}.

\bibitem{Chatrchyan:2011em}
  S.~Chatrchyan {\it et al.}  [CMS Collaboration],
  {\tt 1102.4746 [hep-ex]}.

\bibitem{Rajaraman:2011rw}
  A.~Rajaraman, Z.~Surujon and T.~M.~P.~Tait,
  {\tt 1104.0947 [hep-ph]}.

\bibitem{Feldman:1997qc}
  G.~J.~Feldman and R.~D.~Cousins,
  Phys.\ Rev.\  D {\bf 57} (1998) 3873
  {\tt [physics/9711021]}.


\bibitem{Delaunay:2011gv}
  C.~Delaunay, O.~Gedalia, Y.~Hochberg, G.~Perez and Y.~Soreq,
  {\tt 1103.2297 [hep-ph]}.

\bibitem{Jung:2009pi}
  D.~W.~Jung, P.~Ko, J.~S.~Lee and S.~h.~Nam,
  Phys.\ Lett.\  B {\bf 691} (2010) 238
  {\tt [0912.1105 [hep-ph]]};
  C.~Zhang and S.~Willenbrock,
  Phys.\ Rev.\  D {\bf 83} (2011) 034006
  {\tt [1008.3869 [hep-ph]]};
  C.~Degrande, J.~M.~Gerard, C.~Grojean, F.~Maltoni and G.~Servant,
  {\tt 1010.6304 [hep-ph]}.


\bibitem{Aaltonen:2008hx}
  T.~Aaltonen {\it et al.}  [CDF Collaboration],
  Phys.\ Rev.\ Lett.\  {\bf 102} (2009) 041801
  {\tt [0809.4903 [hep-ex]]}.

\bibitem{Shu:2011au}
  J.~Shu, K.~Wang and G.~Zhu,
  {\tt 1104.0083 [hep-ph]}.

\bibitem{Grinstein:2011yv}
  B.~Grinstein, A.~L.~Kagan, M.~Trott and J.~Zupan,
  {\tt 1102.3374 [hep-ph]}.



\bibitem{Langacker:2000ju}
  P.~Langacker and M.~Plumacher,
  Phys.\ Rev.\  D {\bf 62} (2000) 013006
  {\tt [hep-ph/0001204]}.

\bibitem{Atwood:1996vj}
  D.~Atwood, L.~Reina and A.~Soni,
  Phys.\ Rev.\  D {\bf 55} (1997) 3156
  {\tt [hep-ph/9609279]}.


\bibitem{Buckley:2011vc}
  M.~R.~Buckley, D.~Hooper, J.~Kopp and E.~Neil,
  {\tt 1103.6035 [hep-ph]}.




\end{thebibliography}
\end{document}